\documentclass[prd,reprint,onecolumn,notitlepage,preprintnumbers, showpacs]{revtex4-1}
\usepackage{verbatim}
\usepackage{amsmath, multirow, amssymb, wasysym, gensymb}
\usepackage{graphicx}
\usepackage{units}
\usepackage[utf8]{inputenc}
\begin{document}
\title{Neutrino mass hierarchy and the origin of leptonic flavor mixing from the right-handed sector}
\author{P. Leser}
\email{philipp.leser@tu-dortmund.de}
\affiliation{Fakult\"at f\"ur Physik,
Technische Universit\"at Dortmund, 44221 Dortmund,
Germany}
\author{H. P\"as}
\affiliation{Fakult\"at f\"ur Physik,
Technische Universit\"at Dortmund, 44221 Dortmund,
Germany}
\preprint{TH-DO 11/10}

\begin{abstract}
We consider a neutrino mass model where all leptonic mixing is induced by a heavy Majorana sector through the seesaw type I mechanism, while the Dirac mass matrices are diagonal. Such a pattern occurs naturally in grand unified theories. Constraints on the parameters of the models are considered and it is shown that a normal neutrino mass hierarchy is preferred. The lightest neutrino mass is typically small, leading to nonobservable rates for neutrinoless double beta decay in the normal hierarchy case.
\end{abstract}
\maketitle
\section{Introduction}

In the absence of light right-handed neutrinos, neutrino masses can be generated through the seesaw mechanism. The mass matrix of the neutrinos can then be written as
\begin{align}
	\label{eq:seesawformula}
	M_{\nu}^\text{flv} &= m_\text{D}^T M^{-1} m_\text{D},
\end{align}
where $m_\text{D}$ refers to the matrix of Dirac masses and $M$ is the matrix of Majorana masses.

The mixing of the neutrinos can be described by the Pontecorvo-Maki-Nakagawa-Sakata (PMNS) matrix $V_\text{PMNS}$. At the $3\sigma$ level it can be experimentally determined\cite{GonzalezGarcia:2007ib} to be,
\begin{align}
	\label{eq:pmnsexp}
	V_\text{PMNS} &= \begin{pmatrix}
		[0.77,0.86]	&	[0.50,0.63]	&	[0.00,0.22]\\
		[0.22,0.56]	&	[0.44,0.73]	&	[0.57,0.80]\\
		[0.21,0.55]	&	[0.40,0.71]	&	[0.59,0.83]
	\end{pmatrix}.
\end{align}
Within these bounds, the PMNS matrix is compatible with the tribimaximal (TBM) pattern\cite{Harrison:2002er}:
\begin{align}
	\label{eq:tribimaximal}
	V_\text{PMNS}\approx V_\text{TBM} &= \begin{pmatrix}
		-\frac{2}{\sqrt{6}}	&	\frac{1}{\sqrt{3}}	&	0\\
		\frac{1}{\sqrt{6}}	&	\frac{1}{\sqrt{3}}	&	\frac{1}{\sqrt{2}}\\
		\frac{1}{\sqrt{6}}	&	\frac{1}{\sqrt{3}}	&	-\frac{1}{\sqrt{2}}
	\end{pmatrix}.
\end{align}

\section{Outline of the model}
In this paper we assume that all leptonic mixing originates from a heavy Majorana sector while the Dirac mass matrices of the neutrinos and charged leptons are diagonal. 

A model of this type is quite a natural consequence of $SO(10)$ grand unified theories (GUTs). With all quarks and leptons unified in a $\mathbf{16}$ multiplet of $SO(10)$, the GUT will cause the mass matrices of the quarks and leptons to be very similar. This means that they could at least approximately all be brought into a diagonal form. The experimentally obvious differences between the CKM quark mixing matrix and the PMNS lepton mixing matrix have to be explained by an additional mechanism, which can be the seesaw mechanism used in this paper.

With respect to the seesaw formula Eq.~\ref{eq:seesawformula}, the model investigated here can be written as
\begin{align}
	m_\text{D} &= 
	\begin{pmatrix}
		m_1^\text{D} & 0   & 0\\
		0   & m_2^\text{D} & 0\\
		0   & 0   & m_3^\text{D}
	\end{pmatrix},
\end{align}
and $M \in \mathbb{R}^{3\times 3}$ symmetric and arbitrary. After applying Eq.~\ref{eq:seesawformula}, the mass matrix in the flavor basis is given by
\begin{align}
	M_\text{flv}^\nu&=\frac{1}{\Delta^3}
	\begin{pmatrix}
	 \left[M_{22} M_{33}-\left(M_{23}\right)^2\right] \left(m_1^\text{D}\right)^2
	    & (M_{13} M_{23}-M_{33} M_{12}) m_1^\text{D} m_2^\text{D} & (M_{12} M_{23}-M_{22} M_{13})
	   m_1^\text{D} m_3^\text{D} \\
	 (M_{13} M_{23}-M_{33} M_{12}) m_1^\text{D} m_2^\text{D}
	    & \left[M_{11} M_{33}-\left(M_{13}\right)^2\right] \left(m_2^\text{D}\right)^2 & (M_{12} M_{13}-M_{11} M_{23})
	   m_2^\text{D} m_3^\text{D} \\
	 (M_{12} M_{23}-M_{22} M_{13}) m_1^\text{D} m_3^\text{D}
	    & (M_{12} M_{13}-M_{11} M_{23}) m_2^\text{D} m_3^\text{D} & \left[M_{11} M_{22}-\left(M_{12}\right)^2\right]
	   \left(m_3^\text{D}\right)^2
	\end{pmatrix},
	\label{eq:flavormatrix}
\end{align}
where the common factor of mass dimension three is given by $\Delta^3 = -M_{33} \left(M_{12}\right)^2+2M_{12} M_{13} M_{23}-M_{22} \left(M_{13}\right)^2-M_{11} \left(M_{23}\right)^2+M_{11}M_{22}M_{33}$. It is obvious that the structure of the matrix depends crucially on the differences of the Majorana masses $M_{ij}$.

The $(1,1)$ element of Eq.~\ref{eq:flavormatrix} is the effective mass $m_{\beta\beta}$ observed in neutrinoless double beta decays. In order to make statements about the neutrino mixing angles and squared mass differences, the mass matrix has to be diagonalized using an eigenvalue decomposition, yielding the mixing matrix $U$. We use the ordering scheme of Ref.~\cite{Antusch:2003kp} in which the labels $m_1$ and $m_2$ are assigned to the pair of eigenvalues whose absolute squared mass difference is minimal. Out of these two the eigenvalue whose corresponding eigenvector has the smaller modulus in the first component is labeled $m_2$. The hierarchy of the neutrino masses is then given by the sign of the squared mass differences $\Delta m_{31}^2$ or $\Delta m_{32}^2$. The mixing angles can then be determined using\cite{Ohlsson:1999xb}
\begin{align}
	\theta_{13} &= \arcsin\left(\left|U_{13}\right|\right),\\
	\theta_{12} &= \begin{cases}
	\arctan\left(\frac{\left|U_{12}\right|}{\left|U_{11}\right|}\right) & \text{if~}U_{11}\neq 0\\
	\frac{\pi}{2} & \text{else}
\end{cases},\\
\theta_{23} &= \begin{cases}
\arctan\left(\frac{\left|U_{23}\right|}{\left|U_{33}\right|}\right) & \text{if~}U_{33}\neq 0\\
\frac{\pi}{2} & \text{else}
\end{cases}.
 \end{align}

If the mass matrix in Eq.~\ref{eq:flavormatrix} is supposed to represent neutrino data, it needs to be able to generate neutrino mixing that is close to being tribimaximal\cite{Harrison:2002er}. A general class of flavor space mass matrices that leads to tribimaximal mixing is given by the following pattern\cite{Harrison:2003aw}:
\begin{align}
	\label{eq:tbmpattern}
M_\nu^\text{TBM,flv} &=	\begin{pmatrix}
		x & y & y\\
		y & x+v & y-v\\
		y & y-v & x+v
	\end{pmatrix},
\end{align}
where $x,y$ and $z$ are real numbers. It is useful to compare the entries of this matrix to Eq.~\ref{eq:flavormatrix} for the two important mass hierarchies:
\begin{enumerate}
	\item \textbf{inverted hierarchy:} In this case we approximate an inverted mass hierarchy by two neutrino masses at a higher scale $\tilde{m}$ and one neutrino mass set to zero---i.e., the diagonal mass matrix becomes $\text{diag}(\tilde{m},\tilde{m},0)$. Eq.~\ref{eq:tbmpattern} can then be written as
	\begin{align}
		\label{eq:invmatrix}
		\tilde{m}\cdot
		\begin{pmatrix}
			1 & 0 & 0\\
			0 & \frac{1}{2} & \frac{1}{2}\\
			0 & \frac{1}{2} & \frac{1}{2}		
		\end{pmatrix},
	\end{align}
	which is equivalent to setting $x=1, y=0$ and $v=-\nicefrac{1}{2}$ in Eq.~\ref{eq:tbmpattern}.
	Comparing Eq.~\ref{eq:invmatrix} with the mass matrix of Eq.~\ref{eq:flavormatrix} leads to a set of equations whose solutions determine if an inverted neutrino mass hierarchy is possible in this model:
	\begin{align}
		\frac{1}{\Delta^3}\left(M_{23}^2 - M_{22} M_{33}\right) \left(m_1^\text{D}\right)^2 &= \tilde{m},\nonumber\\
		\frac{1}{\Delta^3}\left(M_{13}^2 - M_{11} M_{33}\right) \left(m_2^\text{D}\right)^2 &= \frac{\tilde{m}}{2},\nonumber\\
		\frac{1}{\Delta^3}\left(M_{12}^2 - M_{11} M_{22}\right) \left(m_3^\text{D}\right)^2 &= \frac{\tilde{m}}{2},\nonumber\\
		\frac{1}{\Delta^3}\left(M_{33} M_{12} - M_{13} M_{23}\right) m_1^\text{D} m_2^\text{D} &= 0,\\
		\frac{1}{\Delta^3}\left(M_{22} M_{13} - M_{12} M_{23}\right) m_1^\text{D} m_3^\text{D} &= 0,\nonumber\\
		\frac{1}{\Delta^3}\left(M_{11} M_{23} - M_{12} M_{13}\right) m_2^\text{D} m_3^\text{D} &= \frac{\tilde{m}}{2}.\nonumber
	\end{align}
	Trying to solve this set of equations immediately leads to a condition
	\begin{align}
		\label{eq:badcondition}
		\tilde{m} &= 0,
	\end{align}
	which means that in this approximation it is not possible to generate an inverted neutrino mass hierarchy. Translated to a realistic scenario where the facts that tribimaximal mixing is only an approximation and that the smaller squared mass difference is not zero are taken into account, one can conclude that in this model the inverted mass hierarchy should be strongly suppressed.
	\item \textbf{normal hierarchy:} A normal neutrino mass hierarchy is approximated by two vanishing neutrino masses and one neutrino mass at a higher scale $\tilde{m}$---i.e., a diagonal mass matrix of $\text{diag}(0,0,\tilde{m})$. This leads to a flavor space mass matrix of the form
	\begin{align}\tilde{m}\cdot
		\begin{pmatrix}
			0 & 0 & 0\\
			0 & \frac{1}{2} & -\frac{1}{2}\\
			0 & -\frac{1}{2} & \frac{1}{2}
		\end{pmatrix}.
	\end{align}
	Again, comparing this matrix with Eq.~\ref{eq:flavormatrix} gives a set of equations. Note that in this case the $(1,1)$ element of the matrix is zero instead of $\tilde{m}$. This eliminates the suppressive condition of Eq.~\ref{eq:badcondition}. The rest of the set of equations is solvable and just restricts the parameter space of the mass matrix.
\end{enumerate}

\section{Numerical analysis}
In this section we present the results of a numerical analysis to determine the general allowed ranges for the following observables:
\begin{enumerate}
	\item[(i)] the neutrinoless double beta decay parameter $m_{\beta\beta}$, given by the $(1,1)$-entry of Eq.~\ref{eq:flavormatrix},
	\item[(ii)] the lightest neutrino mass $m_0$,
	\item[(iii)] the neutrino mixing angle $\theta_{13}$, as it is the mixing angle with the largest experimental uncertainty.
\end{enumerate}
The Dirac masses $m_i^\text{D}$ are mostly responsible for the mass eigenvalues of the neutrinos, while the mixing angles are dominantly determined by the Majorana mass matrix entries $M_{ij}$. 

Because of this, the numerical analysis of each point in the parameter space is performed in two steps:
First, random starting points of the electroweak scale are chosen for the Dirac masses $m_i^\text{D}$, which are then varied. The Majorana parameters $M_{ij}$ are also chosen randomly at a scale of up to $100\times 10^{14}$\,GeV, but are not varied in this step. We have ignored the possibility of $CP$ violation here and assigned real valued numbers to all parameters. A $\chi_M^2$ function for the squared mass differences can be calculated using the diagonalized mass matrix and comparison values from a global fit of all experimental data\cite{GonzalezGarcia:2010er} (the two cited values for $\Delta m_{31}^2$ refer to the two possible mass hierarchies; the uncertainties refer to $1\sigma$ and $3\sigma$ respectively):
\begin{align}
	\left(\Delta m_{21}^2\right)^\text{exp} &= 7.59 \pm 0.20 \begin{pmatrix} +0.61 \\ -0.69\end{pmatrix} \times 10^{-5}\,\text{eV}^2, \\
	 \left(\Delta m_{31}^2\right)^\text{exp} &= \begin{cases}
	-2.36 \pm 0.11 \left(\pm 0.37\right)\times 10^{-3}\,\text{eV}^2\\
	+2.46\pm 0.12 \left(\pm 0.37\right)\times 10^{-3}\,\text{eV}^2
\end{cases}.
\end{align}

This $\chi^2_M$ function is minimized using a multidimensional minimization algorithm\cite{gsl}. At this point, if the minimum is above the threshold value for $\chi^2_M$, the data point is discarded. For the accepted points, the second step consists of calculating the mixing angles\cite{Antusch:2003kp}. The $\chi^2_A$ function for the angles is then analyzed and compared to the data obtained from the global fit of experimental results\cite{GonzalezGarcia:2010er}:
\begin{align}
	\theta_{12}^\text{exp} &= 34.4\pm 1.0 \begin{pmatrix} +3.2\\ -2.9 \end{pmatrix}\degree, & 	\theta_{23}^\text{exp} &= 42.8 \begin{array}{c}
		+4.7 \\ -2.9
	\end{array} \begin{pmatrix} +10.7\\ -7.3 \end{pmatrix}\degree, & 	\theta_{13}^\text{exp} &= 5.6 \begin{array}{c} +3.0\\ -2.7 \end{array}\left(\leq 12.5\right)\degree.
\end{align}
Note that the cited value for $\theta_{13}$ is based on a global fit including the other angles and mass observables as no direct measurement of the angle exists as of now.

If the $\chi^2_A$ value is below the threshold value (for the number of degrees of freedom considered here, it is $\chi^2_A<16.8$ for a significance of 0.99), the data point as well as the values of the observables are recorded.

The scan covering roughly $16\times 10^6$ data points finds 177256 acceptable samples that lead to a normal neutrino mass hierarchy and 3596 samples with an inverted neutrino mass hierarchy. This meets the expectation of the inverted mass hierarchy being suppressed in the model. The best fit point lies in the regime of normal mass ordering with $\chi^2_A=0.005$. The parameters leading to this best fit are
\begin{align}
	  M_{11} &= 44\times 10^{13}\,\text{GeV}
	& M_{22} &= 24\times 10^{13}\,\text{GeV}
	& M_{33} &= 89\times 10^{13}\,\text{GeV}\nonumber\\
	  M_{12} &= 97\times 10^{13}\,\text{GeV}
	& M_{13} &= 97\times 10^{13}\,\text{GeV}
	& M_{23} &= 21\times 10^{13}\,\text{GeV}\\
	 m_1^\text{D} &= 46\,\text{GeV}
	&m_2^\text{D} &= 140\,\text{GeV}
	&m_3^\text{D} &= 130\,\text{GeV}\nonumber  
\end{align}

\begin{figure}[p]
	\centering
		\includegraphics[width=8.6cm]{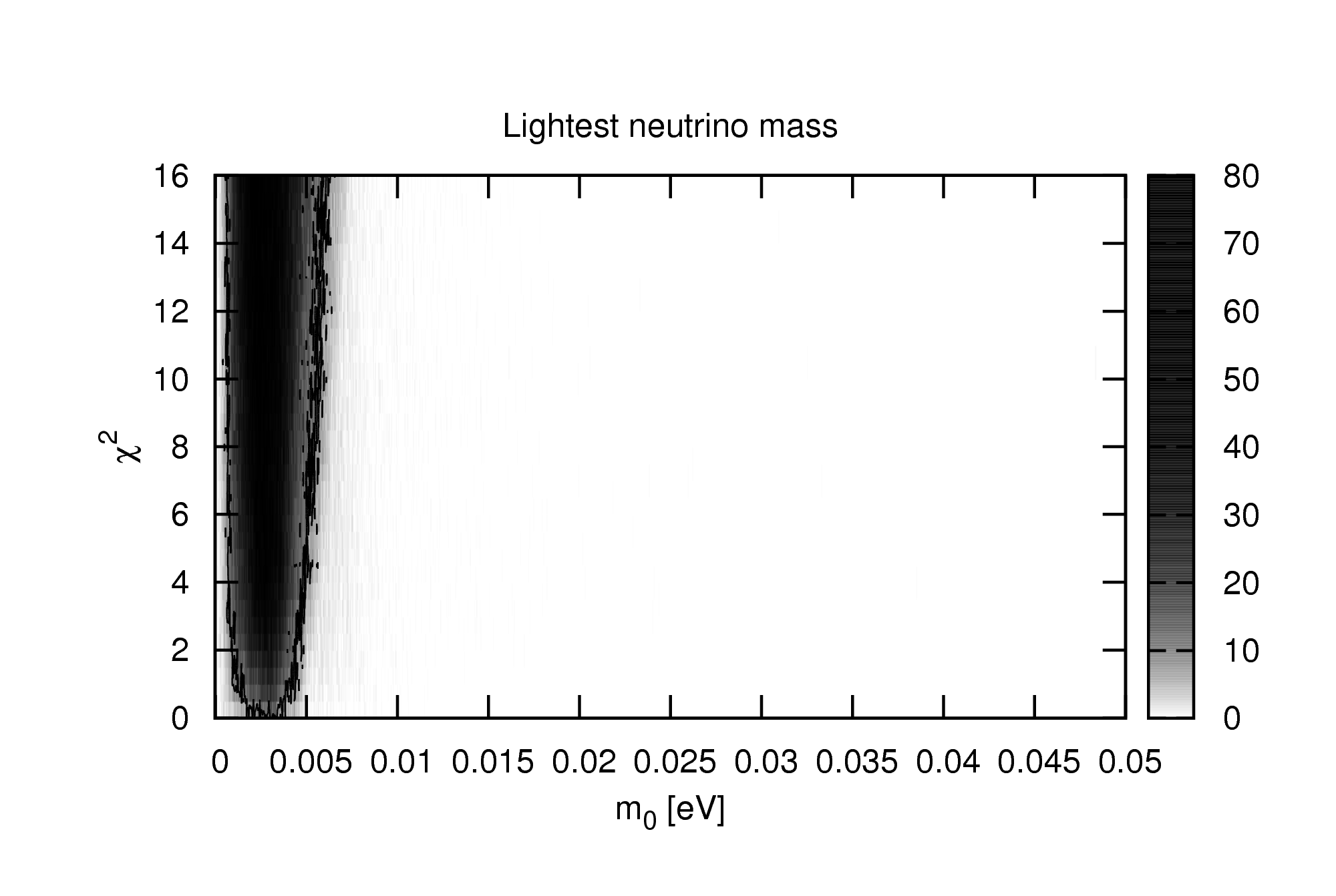}
	\caption{The lightest neutrino mass $m_0$ for all models with $\chi^2_A < 16.8$ for the case of a normal neutrino mass hierarchy. The shade is proportional to the number of hits that lie in the shaded bin. The plot has been divided into 1668 bins on the $x$ axis and 35 bins on the $y$ axis. Outside of the boundary line the hit density is less than 10 hits per bin.}
	\label{fig:klnormlightest}
\end{figure}

\begin{figure}[htbp]
	\centering
		\includegraphics[width=8.6cm]{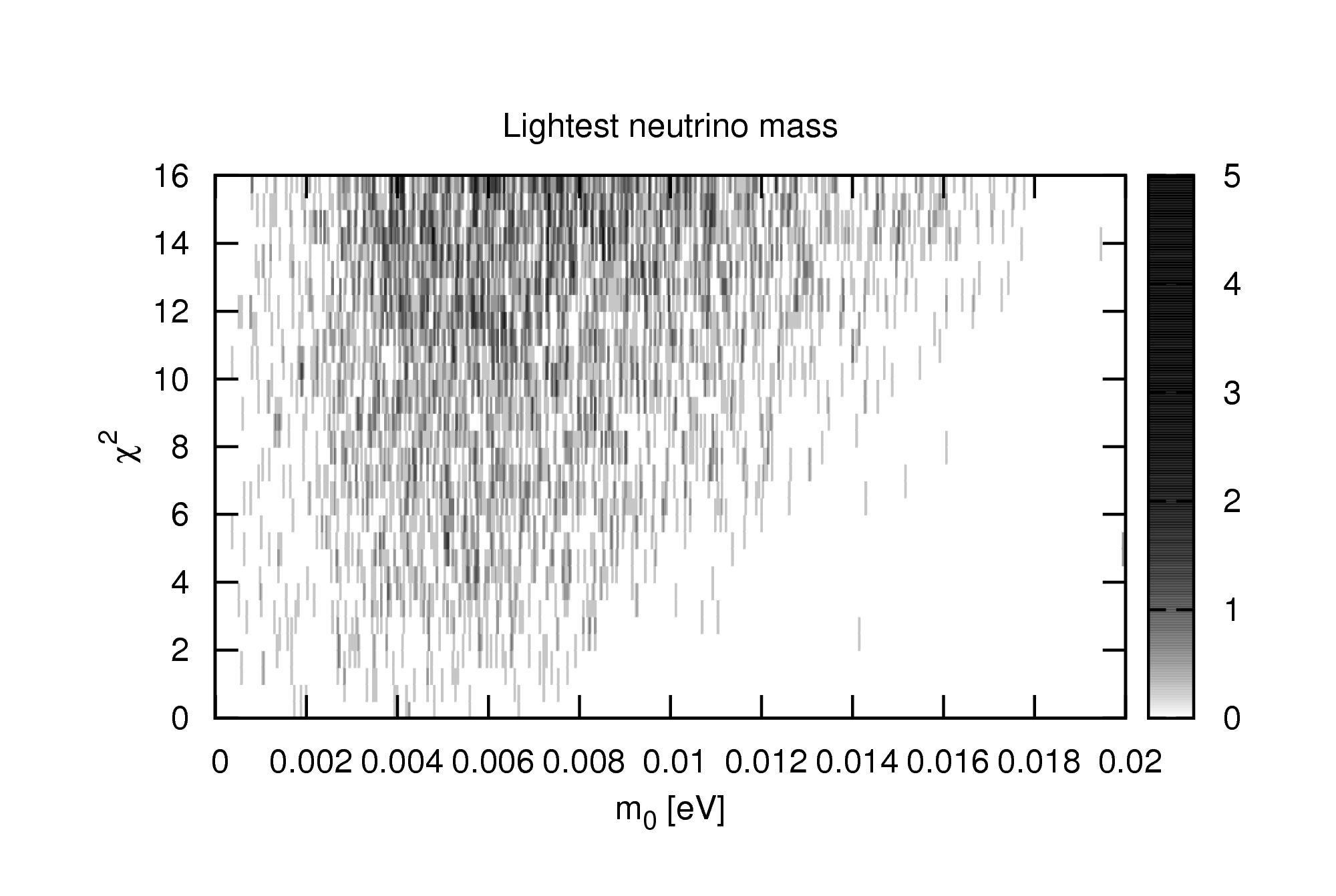}
	\caption{The lightest neutrino mass $m_0$ for all models with $\chi^2_A < 16.8$ for the case of an inverted neutrino mass hierarchy.  The shade is proportional to the number of models that lie in the shaded region. The plot has been divided into 689 bins for $x$ axis and $35$ bins for the $y$ axis.}
	\label{fig:klinvlightest}
\end{figure}

For all acceptable points, the values for the lightest neutrino mass $m_0$ are displayed in Fig.~\ref{fig:klnormlightest} for a normal mass hierarchy and in Fig.~\ref{fig:klinvlightest} for an inverted neutrino mass hierarchy. 

For both hierarchies, the best fit lightest neutrino mass $m_0$ is below $0.005$\,eV, although larger values of up to $0.02$\,eV are still acceptable for the inverted hierarchy case (Fig.~\ref{fig:klinvlightest}). Both values lead to summed neutrino masses $M_\nu=\sum m_i$ below the current bound of roughly $M_\nu\approx 0.5$\,eV\cite{Steidl:2009hx,Elgaroy:2002bi,Tegmark:2006az,Hannestad:2006mi,Spergel:2006hy,Seljak:2006bg,Allen:2003pta,   Hannestad:2006as,Hannestad:2007cp,Lesgourgues:2004ps}.

As the contribution from $m_0$ is negligible and $\theta_{13}$ is small, the neutrinoless double beta decay observables for the cases of normal or inverted hierarchies are given by (see e.g.~Refs.~\cite{KlapdorKleingrothaus:2000gr, Pas:2001nd, Petcov:2004wz,Lindner:2005kr} and the references therein)
\begin{align}
	m_{\beta\beta} &\approx
	\begin{cases}
		\sqrt{\Delta m^2_{12}}\sin^2\left(\theta_{12}\right) & \text{for normal hierarchy}\\
		\sqrt{\Delta m^2_{\text{23}}}\quad\text{resp.}\quad\sqrt{\Delta m^2_{\text{23}}} \cos\left(2\theta_{12}\right) & \text{for inverted hierarchy} 
	\end{cases},
\end{align}
where $\Delta m_{12}\ll\Delta m_{23}$ and $CP$ conservation has been assumed. The two values given for the case of an inverted mass hierarchy stem from the sign ambiguity of the $\pm \sin^2\left(\theta_{12}\right)$ term in the sum of masses. Using these formulas, the squared mass differences and mixing angles given in the global fit of Ref.~\cite{GonzalezGarcia:2010er} lead to $m_{\beta\beta}^\text{normal} \approx 0.003$\,eV for the case of a normal mass hierarchy. The two possible values for the inverted mass hierarchy case are $m_{\beta\beta}^\text{inv/a} \approx 0.02$\,eV and $m_{\beta\beta}^\text{inv/b} \approx 0.05$\,eV. As the model discussed in this article reproduces the mixing angles and mass parameters of the global fit, the predictions for the mass parameter of the neutrinoless double beta decay $m_{\beta\beta}$ in that case also hold. 

All angles can be fitted to the experimental data\cite{GonzalezGarcia:2010er}. As there is no direct measurement for $\theta_{13}$, it is instructive to also take a look at the case where $\theta_{13}$ is not part of the fit. In that case, only the squared mass differences and the angles $\theta_{12}$ and $\theta_{23}$ are fitted. The allowed ranges for the angle $\theta_{13}$ can then be seen as a prediction of the model. In the case of a normal mass hierarchy, there is a clear preference for $\theta_{13}\approx 0$, although values up to and beyond the bound of $\theta_{13}^\text{exp}<0.15$\cite{Nakamura:2010zzi} are certainly possible. In the case of an inverted mass hierarchy, the preferred value of $\theta_{13}$ is clearly close to a maximal mixing angle, although there are results that satisfy the experimental bound.

Consequently for the preferred case of normal mass hierarchies the model typically predicts small values for $\theta_{13}$ outside of the sensitivity range of Double Chooz\cite{Ardellier:2006mn,Novella:2011tk}, Daya Bay\cite{Guo:2007ug,Lin:2011pb} and Reno\cite{Ahn:2010vy,Kim:2010zz} (see also Refs.~\cite{Huber:2009cw, Lasserre:2005qw}). For the less favored inverted hierarchy, however, upcoming reactor experiments have good prospects for measuring $\theta_{13}$.

Recently, the T2K experiment\cite{Itow:2001ee} has observed indications of $\nu_\mu\to\nu_e$ appearance\cite{Abe:2011sj}. The observed number of events deviates from the expectation compatible with $\sin^2 2\theta_{13}=0$ with a signifiance of $2.5\sigma$. According to Ref.~\cite{Abe:2011sj} this leads to an allowed range of $0.03(0.04) < \sin^2 2\theta_{13} < 0.28(0.34)$ for a normal (inverted) mass hierarchy. Even though this model prefers a large value for $\theta_{13}$ in the inverted hierarchy case, the predictions remain compatible with the T2K bounds within the allowed $\chi^2$ range for both normal and inverted mass hierarchies.

\section{Conclusions}
In this paper, a generic model based on a seesaw type 1 mechanism with diagonal Dirac mass matrices for both the charged leptons and the neutrinos has been considered. All contributions to the observed neutrino mixings originate from the heavy Majorana masses through a generic Majorana mass matrix that allows for off-diagonal components. It has been shown that models of that kind---which are well-motivated by GUTs---generate a small mixing angle $\theta_{13}$ naturally and that a normal neutrino mass hierarchy is preferred.

\end{document}